\documentclass[letterpaper,12pt]{article}   %% LaTeX 2e (preferred)
\usepackage{osajnl2} %% do not use with REVTeX4

\begin{document}

\title{Profiling of micrometer sized laser beams in restricted volumes}

%% For REVTeX it is possible to automate superscript and e-mail callouts with the superscriptaddress option; see REVTeX4 documentation.

\author{{Yevhen Miroshnychenko,$^{1}$ Otto Nielsen,$^1$ Aske Thorsen,$^1$ and Michael Drewsen$^{1,*}$}
\address{$^1$QUANTOP, Danish National Research Foundation Center for Quantum Optics, Department of Physics and Astronomy, University of Aarhus, DK-8000 Aarhus C, Denmark}
\address{$^*$Corresponding author:drewsen@phys.au.dk}
}

\begin{abstract} We present a method for determining the 3D intensity distribution of directed laser radiation with micrometer resolution in restricted volumes. Our method is based on in-coupling and guiding properties of optical fibers, with the current version requiring only few hundred micrometers. We characterize the  performance of the method and experimentally demonstrate profiling of micrometer sized laser beams. We discuss the limiting factors and routes towards a further increase of the resolution and beam profiling in even more restricted volumes. Finally, as an application example, we present profiling of laser beams inside a micro ion trap with integrated optical fibers.
\end{abstract}

%120.3940 Instrumentation, measurement, and metrology : Metrology
%140.3295 Lasers and laser optics : Laser beam characterization
%060.2370   Fiber optics sensors
%060.4005   Microstructured fibers
%130.0130   Integrated optics   
%020.0020   Atomic and molecular physics   
%180.5655   Raman microscopy

\maketitle %% null function with osajnl.sty
\section{Introduction}
A method for measuring 3D intensity distribution of directed laser radiation in a volume-restricted environment may be of importance in many different laser light based fields of industry and science. For example, it would allow for diagnostics of optical communication devices (e.g. fiber multiplexers), including fast diagnostics during manufacturing, final quality control and service of such devices. Micro beam profiling may as well be an important tool in connection with optical data storage, where extremely focused and high quality laser beams are needed and quick high quality beam diagnostics in restricted volumes is essential. Another example is multi-photon Raman spectroscopy \cite{Nibler87}, where several focused laser beams with required sizes should be carefully overlapped in space. 
Furthermore, in the fast growing field of experimental quantum optics, miniaturization of atomic particle traps with integrated optical elements, e.g. micro optical resonators and optical waveguides, plays one of the key roles \cite{Colombe07,Kim09}. In such experimental configurations it is desirable to know to a high degree the characteristics of the beams emerging after the integrated optics. Although there are many commercially available solutions for beam profiling, none of them so far allowed to perform the beam characterization \textit{in-situ} in restricted environments to check the quality of the assembly and characterize the optical part of the integrated system. 

Existing directed laser radiation profiling techniques can be divided into three classes. The first class is based on direct projection of the whole laser beam onto a two-dimensional array of photo sensitive pixels such as photo-diodes or a CCD chip for the following profile analysis \cite{Knudson83}. The second class is based on a controlled shading of the laser radiation and recording the corresponding intensity reduction of the uncut beam. The shading can be done using a knife edge \cite{Arnaud71}, a translating slit \cite{Giles84}, a pinhole \cite{Shayler78,Brannon75}, a rotating mirror \cite{Wang84} or a wire \cite{Lim82}. The third class is based on spectroscopic \cite{Granneman75,Sorscher80} or thermographic \cite{Rose86,Baba86} methods. The first two classes in the present form require bulky detectors in the vicinity of the measuring point both in the radial and axial extensions. The latter class, additionally, is based on properties of a medium around the measured laser beam and, therefore, is applicable only for special types of lasers.

\begin{figure}[htbp]
  \centering
\includegraphics{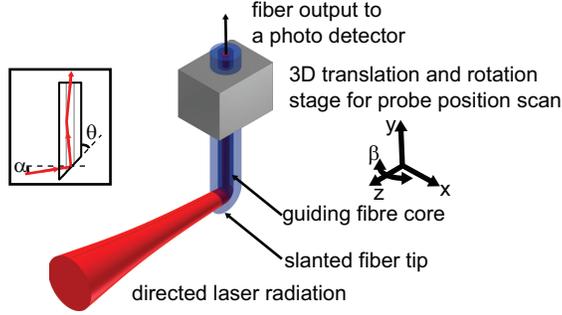}
\caption{\label{fig:principle} %(Color online)
Principle of operation of the laser beam probe. 
%A narrow spatially filtering detector head in the form of a tip of an optical fiber picks up a part of the directed laser radiation intensity at a certain position (see the inset). The optical fiber itself guides the light to a photo detector, which allows for measuring a signal proportional to the directed laser radiation intensity at the position of the fiber tip at any convenient location.
}
\end{figure}

Here we present a beam profiling method allowing micrometer resolution and requiring available "working volumes" of at most only several hundred micrometers across around the measuring position, which is enough to introduce our detector head based on the optical fiber. Additionally, the same method can be used to find the intersection point of several directed laser radiation sources even in volumes with restricted access to the laser beams. Finally, the micrometer resolution enables to use the devise for pointing stability measurements of directed laser radiation.

\section{Principle of operation}
The scanning part of our probe is a slanted tip of an optical fiber, see Fig.~\ref{fig:principle}. Directed laser radiation enters the fiber from the side. The operation of our beam probe is based on the interplay of three physical effects. First, the polished side of the optical fiber acts as a mirror to guide the light into the mode of the optical fiber, see the inset of Fig.~\ref{fig:principle}. %The role of the optical fiber is twofold. 
Second, at the fiber tip interface, laser radiation from only a micrometer-size area is coupled into the guiding mode of the fiber. Finally, the coupled light is guided to a remote photo-detector. This results in an electrical signal proportional to the laser radiation intensity at the position of the fiber tip. Scanning the position of the probe tip, one can reconstruct the 3D profile of the laser radiation pattern.
\begin{figure}[htbp]
  \centering
\includegraphics{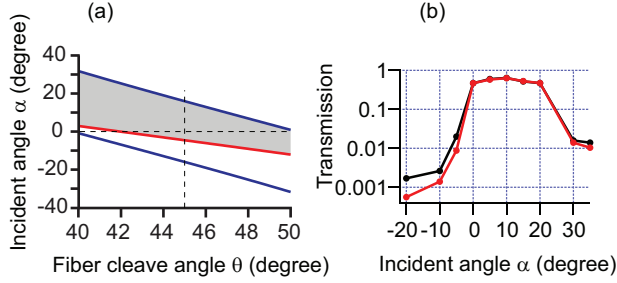}
\caption{\label{fig:polarizationDependence} %(Color online)
%(a) Acceptance_range_calculation.m
%(b) Results summary\MM fiber different beamsizes.pxp
Light coupling into a fiber probe through a slanted tip. (a) Calculated high coupling region (gray) as the function of the fiber polishing angle $\theta$ and the incident angle $\alpha$. (b) Measured transmission of a $22~\mu$m-waist beam at $\lambda=866$~nm through a fiber probe with $\theta=45^{\circ}$ for the vertical  (red) %, i.e. along the fiber core,
and the horizontal light polarizations (black). 
%(b) Measured beam profiles for different beam sizes.
% (multimode fiber GIF62.5)
}
\end{figure}

The efficient side in-coupling into the fiber probe can be done either by using reflection coating of the polished surface or relying on total internal reflection (TIR) on the glass-air (vacuum) interface. The last method does not require extra coating, but only polishing at the proper angle $\theta$. If the beam in-coupling angle $\alpha$, see the inset of Fig.~\ref{fig:principle}, is larger than the critical angle for total internal reflection, then just polishing of the fiber is enough to ensure reflection and does not require expensive reflection coating. Ideally, the reflecting side should be at exactly $\theta=45^{\circ}$, but the TIR condition is not satisfied at this angle for all silica optical fibers in air at typical wavelengths. Additionally, since at the boarder of the TIR condition the reflection is polarization sensitive, it is desirable to work in a deep TIR regime to avoid sensitivity to slight misalignments. The area above the red line in Fig.~\ref{fig:polarizationDependence}a shows the calculated TIR region for a multimode fiber GIF62.5 at 866~nm. The reflected light should at the same time be within the numerical aperture of the optical fiber in order to be coupled into the fiber guiding mode, see the area between the blue lines in Fig.~\ref{fig:polarizationDependence}a. The nonzero overlap between these areas around $\alpha=0^{\circ}$ allows the operation of our probe. Therefore, as far as we reach the TIR regime, the spatial selectivity of our probe is polarization insensitive and is mainly defined by the in-coupling efficiency into an optical fiber. 
\begin{figure}[htbp]
  \centering
\includegraphics{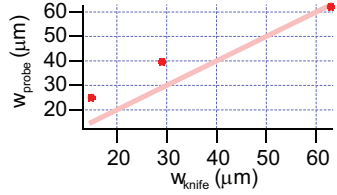}
\caption{\label{fig:MMcalibration} %(Color online)
% GHMMCalibration from ScanXRazorbladeFromASKE.pxp
Beam waists measured with the multimode fiber probe along the $x$-axis, compared to the knife edge method. %The solid line with the slope one corresponds to a probe with infinitely narrow TF.
}
\end{figure}

Generally, profiles recorded with this technique are convolutions of a 
beam shape with a transmission function (TF) of the scanning probe. In the simplest case of coupling light from a free propagating Gaussian beam into a normal cut single mode fiber, the TF of the probe is well approximated by a Gaussian \cite{Marcuse77}.
Assuming that the slanted end of the fiber works just as a mirror (neglecting the lensing effect of the curved entrance surface), the corresponding transmission of the fiber probe as the function of the displacement $d$ in the $x-y$-plane at the beam waist is an overlap between the incoming beam and the fiber guiding mode: $T\left( d\right)=\frac{4 w_b^2 w_f^2}{(w_b^2 +w_f^2)^2}\exp \left(- \frac{2d^2}{w_b^2 +w_f^2} \right),$
where $w_f$ is the effective width of the fiber probe TF, and $w_b$ is the beam size along this scan direction (all beam sizes are given as a $1/e^2$-radius). The resulting profile has a Gaussian shape with the effective width $w_p=\left(w_b^2 +w_f^2\right)^{1/2}$. Consequently, the smaller is $w_f$, relative to the beam size, the tinier is the contribution from the convolution.

 At the same time, the amount of transmitted light drops with the reduction of $w_f$. Much higher coupling efficiency can be reached with a multimode fiber. In this case the efficiency is given by an overlap between the incoming beam and all guiding modes of the fiber. Generally, it is quite difficult to analytically account for all the guiding modes of a multimode fiber. Therefore, it is more practical to directly calibrate the width measured with such a fiber probe using some independent method, for example a knife edge, see Fig.~\ref{fig:MMcalibration}. Such calibration once performed, can be used in all further measurements with this probe.

%measure directly the size of the beam with the probe and compare to the measurement using some independent method, for example using a razor blade, see Fig.~\ref{fig:power}c. Such curve can be directly used as a calibration of the transmission function of the scanning probe...

\section{Characterization of the probe}
Depending on the laser beam sizes, fiber beam profilers can be based on fibers with different core sizes. For beam sizes larger than ca. $20~\mu$m one can choose multimode fibers with the core diameters of several tens of micrometers. Whereas for smaller beam sizes one can choose single mode fibers. We have manufactured and characterized several of such fiber probes.

We first characterized the properties of a fiber beam profiler based on a multimode fiber GIF62.5 with  $\theta=45^\circ$, which corresponds to the vertical dashed line in Fig.~\ref{fig:polarizationDependence}a. 
Figure~\ref{fig:polarizationDependence}b shows a measured transmission as the function of the in-coupling angle $\alpha$ for two light polarizations. We clearly observe strong polarization dependence for $\alpha < -5^{\circ}$ as we approach the TIR edge and polarization independence for higher angles. 

Second, we manufactured a probe based on the same fiber but with $\theta=46^\circ$, allowing us to reliably profile beams at an angle $\alpha=0^{\circ}$. We used this probe to measure radial profiles of beams with different waists. These beams were independently characterized using the knife edge method to perform the calibration of the probe, see Fig.~\ref{fig:MMcalibration}. Since the measured beam size is a convolution of the original beam shape with the TF, for beam sizes comparable with the fiber core radius of $31.25~\mu$m we observe deviation from the curve with the slope one, corresponding to a probe with infinitely narrow TF. At the same time, for larger beam sizes this deviation is negligible. 
\begin{figure}[htbp]
  \centering
\includegraphics{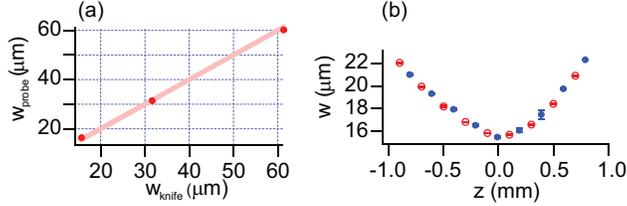}
\caption{\label{fig:SMCalibration} %(Color online)
%a) GBVbisNewShifted from Beams charachterization_analysis.pxp
%b) GVSMCalibration from Beams charachterization 30 um beam C.pxp
Beam sizes measured with the single mode fiber probe along the $y$-direction, compared to the knife edge method. a) Waists measured for different beams. %All the points in the measured range do not significantly deviate from the line with the slope one. %, corresponding to a probe with infinitely narrow TF
b) Beam sizes measured for a beam with $w_b=15.7~\mu$m using the knife edge method (blue dots) and the fiber probe (red circles).% Note that the centers of the curves were superimposed, since the absolute zeros of the two methods are different.
}
\end{figure}

Third, to demonstrate the operation of a beam profiler based on a single mode fiber, we manufactured a probe with $\theta=46^{\circ}$ based on a fiber SM800-5.6-125 with the effective core radius $w_f=2.9~\mu$m at 866~nm. For this small $w_f$ the contribution from the convolution is below 10~\% even for beams as small as $w_b=6.5~\mu$m. Figure~\ref{fig:SMCalibration}a shows the beam waists measured with this probe, compared to the knife edge method. Therefore we do not see now in the measured range any significant deviations from the line with slope one. Figure~\ref{fig:SMCalibration}b shows beam sizes along the $y$-direction as the function of the axial position along the beam. The measured shape is in a perfect agreement with the reference measurement using the knife edge method. For this beam we measure with the fiber probe waists $w_p^x=15.0(4)~\mu$m and $w_p^y=15.5(4)~\mu$m for the scans along the $x$- and $y$-directions. These waists agree quite well with corresponding knife edge measurements $w_b^x=15.7(4)~\mu$m and $w_b^y=15.7(4)~\mu$m, respectively. This demonstrates the feasibility of 3D profiling of micrometer sized beams with single mode fiber probes.

Finally, we experimentally verified the possibility to couple light into the probe not only directly at $\beta=0^{\circ}$, see Fig.~\ref{fig:principle}, but at different angles. The maximum measured transmission through the multimode fiber probe is still $20~\%$ at $\beta=\pm 45^{\circ}$, relative to the direct coupling. %see 04/11/08 from the Labbook
This result opens the possibility to use such a probe for free space beam overlap.

\section{Conclusion and outlook}
\begin{figure}[htbp]
  \centering
\includegraphics{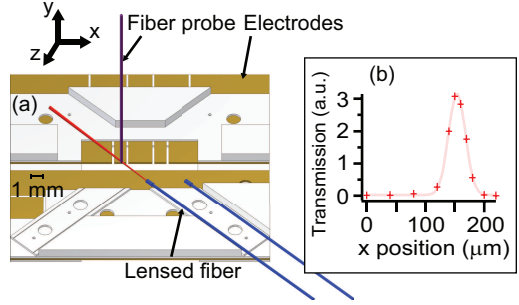}
\caption{\label{fig:trapexample} %(Color online)
% (b) Yevhens paper editing_Insitu IR beam profile from_ LensedBeamSizeChek.pxp
Probing beams from optical fibers integrated into a micro ion trap. (a) Schematic of a partly assembled Paul trap. The fibers are integrated between the layers of electrodes. The upper electrodes are not shown. (b) Example of a beam profile measured in-situ inside the ion trap.}
\end{figure}

We have used our fiber probe for \textit{in-situ} characterization of beams from two single mode lensed fibers inside a micro ion trap, see Fig.~\ref{fig:trapexample}a. Using this method we aligned the fibers so that both beams are crossing at the center of the ion trapping region and characterized the beams afterwards. Figure~\ref{fig:trapexample}b shows an example of a 1D scan through one of the beams at $\lambda=397$~nm.  

We have presented and successfully experimentally tested a beam profiling method with micrometer resolution capable of working  in spatially restricted environments. %We tested the method for free propagating beams and found the results in a good agreement with an independent knife edge measurements. %The test measurements for free propagating beams are in good agreement with the independent knife edge measurements. 
Our technique requires available "working volumes" of at most only several hundred micrometers across in the vicinity of the measuring position, which is enough to introduce our detector head. A further reduction of the working volume and increase of the resolution can be achieved by radial etching, tapering of the fiber or structural coating of the in-coupling tip.

\section*{Acknowledgments}
The authors want to thank P. Balling for the discussions on the early stage of the project. We acknowledge financial support from the Danish National Research Foundation's Center for Quantum Optics QUANTOP (JNR. 02-512-44), the EU Consortium Project MICROTRAP (cn 517675-2) and Y.M. was supported by a personal Marie Curie research grant (cn 236417).

\providecommand{\noopsort}[1]{}\providecommand{\singleletter}[1]{#1}%

%\bibliographystyle{osajnl}
%\bibliography{PF_references}

\begin{thebibliography}{10}
\newcommand{\enquote}[1]{``#1''}

\bibitem{Nibler87}
J.~W. Nibler and J.~J. Yang, Ann. Rev. Phys. Chem. \textbf{38}, 349 (1987).

\bibitem{Colombe07}
Y.~Colombe, T.~Steinmetz, G.~Dubois, F.~Linke, D.~Hunger, and J.~Reichel,
  Nature (London) \textbf{450}, 272 (2007).

\bibitem{Kim09}
J.~Kim and C.~Kim, Quant. Inf. Comput. \textbf{9}, 181 (2009).

\bibitem{Knudson83}
J.~Knudtson and K.~Ratzlaff, Rev. Sci. Instr. \textbf{54}, 856 (1983).

\bibitem{Arnaud71}
J.~Arnaud et.al., Appl. Opt. \textbf{10}, 2775 (1971).

\bibitem{Giles84}
M.~Giles and E.~Kim, SPIE Conference on ¯ber Optics: Short-haul and Long-haul
  Measurements and Applications \textbf{500}, 67 (1984).

\bibitem{Shayler78}
P.~Shayler, Appl. Opt. \textbf{17}, 2673 (1978).

\bibitem{Brannon75}
P.~Brannon et.al., J. of Appl. Phys. \textbf{46}, 3567 (1975).

\bibitem{Wang84}
C.~Wang, Appl. Opt. \textbf{23}, 1399 (1984).

\bibitem{Lim82}
G.~Lim and W.~Steen, Optics and Laser Technology \textbf{14}, 149 (1982).

\bibitem{Granneman75}
E.~Granneman et.al., Rev. Sci. Instr. \textbf{46}, 332 (1975).

\bibitem{Sorscher80}
S.~Sorscher and M.~Klein, Rev. Sci. Instr. \textbf{51}, 98 (1980).

\bibitem{Rose86}
A.~Rose et.al., Appl. Opt. \textbf{25}, 1738 (1986).

\bibitem{Baba86}
T.~Baba et.al., Rev. Sci. Instr. \textbf{57}, 2739 (1986).

\bibitem{Marcuse77}
D.~Marcuse, The Bell System Technical Journal \textbf{56}, 703 (77).

\end{thebibliography}
\end{document}